\RequirePackage{fix-cm}
\documentclass[twocolumn,epjc3]{svjour3}  
\smartqed  % flush right qed marks, e.g. at end of proof
\RequirePackage{graphicx}
\usepackage{lipsum}
\usepackage{amsmath}
\usepackage{amssymb}
\usepackage{multicol}
\usepackage{cuted}
\usepackage{subcaption}

\RequirePackage[numbers,sort&compress]{natbib}
\RequirePackage[colorlinks,citecolor=blue,urlcolor=blue,linkcolor=blue]{hyperref}
\journalname{Eur. Phys. J. C}
\begin{document}

\title{A note on Gravitational radiation in generalized Brans–Dicke theory: compact binary systems}

\author{Diego S. Jesus\thanksref{e1,addr1}
        \and
        Hermano Velten\thanksref{e2,addr2}
        \and
        Júnior D. Toniato\thanksref{e3,addr3}
}

\thankstext{e1}{e-mail: \href{mailto:diego@...}{diego.s.jesus@edu.ufes.br}}
\thankstext{e2}{e-mail: \href{mailto:hermano.velten@edu.ufop.br}{hermano.velten@edu.ufop.br}}
\thankstext{e3}{e-mail: \href{mailto:junior.toniato@ufes.br}{junior.toniato@ufes.br}}

\institute{Depto.  Física \& Núcleo Cosmo-Ufes, Universidade Federal do Espírito Santo, Campus Vitória, Brazil. \label{addr1}
           \and
           Departamento de Física, Universidade Federal de Ouro Preto (UFOP), Ouro Preto, Brazil. \label{addr2}
           \and
           Depto. Química e Física \& Núcleo Cosmo-Ufes, Universidade Federal do Espírito Santo, Campus Alegre, Brazil\label{addr3}
}

\date{Received: date / Accepted: date}

\maketitle

%\begin{abstract}
%Insert your abstract here. Include keywords, PACS and mathematical
%subject classification numbers as needed.
%\end{abstract}

%\section{Introduction}

In a recent article, the authors of Ref. \cite{Mahmoudi:2024rga} study the gravitational radiation generated by compact binaries in a Brans-Dicke-$\!{f(R)}$ (BD$f$) theory based on the following action
\begin{eqnarray}
A = \frac{1}{16\pi} \int \sqrt{-g} \left[ \phi f(R) - \frac{\omega_0}{\phi} \partial_\mu \phi \, \partial^\mu \phi \right] d^4x + A_m.
\label{action}
\end{eqnarray}
It is assumed that the scalar field $\phi$ is massless and has a constant coupling parameter $\omega_0\equiv cte$. Also, $A_m\equiv A_m(g_{\mu\nu},\Psi_{m})$ denotes the matter action and $\Psi_m$ represents the matter fields.
%The original Brans-Dicke action is generalized by introducing a general $f(R)$ function replacing the Ricci scalar ($R$).
This class of theories is also known as ``first generation'' scalar-tensor theories. 

After varying the above action, one finds the trace of the field equation
\begin{equation}
    f_R R - 2 f(R) + \frac{3 \Box (\phi f_R)}{\phi} + \frac{\omega_0}{\phi^2} \partial_\mu \phi \partial^\mu \phi = \frac{8 \pi T}{\phi},
\end{equation}
as well as an equation for the evolution of the scalar field
\begin{align}
\Box \phi - \frac{\partial_\mu \phi \partial^\mu \phi}{4 \phi} = \frac{1}{4 \omega_0} \Big[ 8 \pi \Big(T &- 4 \phi \tfrac{dT}{d \phi}\Big) \notag\\
&- \phi R f_R - 3 \Box (\phi f_R) \Big],
\end{align}
where $f_R\equiv \partial f/ \partial R$. 
%The weak field limit of the above equations is obtained perturbing the flat space-time, i.e introducing the tensor field $h_{\mu \nu}$, as $g_{\mu \nu} = \eta_{\mu \nu} + h_{\mu \nu},$ where $\eta_{\mu \nu}$ is the Minkowski metric, and $h_{\mu \nu}$ denotes a small deviation with respect to the flat spacetime, i.e. $|h_{\mu \nu}| \ll 1$ with trace $h = \eta^{\mu \nu} h_{\mu \nu}$.
As usually understood, $f(R)$ theories can be recast as massive scalar-tensor models through the identification $f_R = \Phi$. Thus, the BD$f$ theory is essentially a metric theory of gravity with two more scalar fields: the usual massless BD field $\phi$, and the effectively geometrical massive field $\Phi$.
%According to the above equations, the GBD theory is interpreted as a two-scalar-field theory, i.e., the usual BD field and the effectively geometrical field $f_R = \Phi$.

The parameter $\omega_{0}$ controls the influence of the massless scalar field $\phi$ on space-time curvature. For the model exposed in \eqref{action}, low values of $\omega_{0}$ indicate significant deviations from a single $f(R)$ theory. The influence of the geometric scalar field $\Phi$ is encoded in its mass $m_f$, as it is called in Ref. \cite{Mahmoudi:2024rga}. A small mass value would strengthen the effects of nonlinear Ricci terms in the action, while the limit $m_f\rightarrow\infty$ recovers the massless BD theory.

The most stringent constraint on $\omega_{0}$ comes from Solar System tests. The Cassini mission, by precisely measuring the time delay of radio signals passing near the Sun, established for small mass cases of the BD theories: $\omega_{0} \gtrsim 40000$ with $m_\phi\lesssim2.5\times 10^{-29}$ GeV \cite{Alsing:2011er}.\footnote{The large mass case (with $\omega_0$ still inversely proportional to $m_\phi$) have been recently constrained by Mercury's perihelion shift \cite{Alves:2023cuo}. For $m_\phi\gtrsim 4.6\times 10^{-26}$ GeV one has $\omega_0\gtrsim 20000$, while for $m_\phi\gtrsim 10^{-25}$ GeV, there is no bound for $\omega_0$.}
It is worth noting that this limit is based on generalized BD models, where the scalar field $\phi$ has a mass $m_\phi$, and it does not consider any influence of other fields.
Thus, the parameters $m_f$ and $m_\phi$ do not represent the same physical quantity, and they relate to $\omega_0$ distinctly.

The main result of Ref. \cite{Mahmoudi:2024rga} is a lower bound $\omega_0 > 6.09723 \times 10^6$, for a massless BD field with geometrical field mass smaller than $10^{-29}$ GeV, being this result ``two orders of magnitude stricter than those derived from solar system data'', as stated by the authors. This would be quite a relevant result, but it should be more carefully analyzed, and this is our first important remark in this note. The strongest constraint obtained is highly dependent on the influence of the massive field too, reason why one should not expect also that the parameter $\omega_0$ in this model has the same role as the traditional BD parameter. 

\begin{figure*}
\centering
\begin{subfigure}{0.32\textwidth}
\centering
\includegraphics[width=\linewidth]{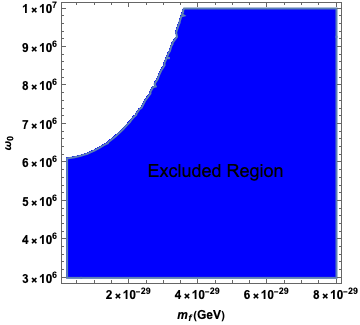}
\caption{}
\label{fig:MVplot}
\end{subfigure}
\begin{subfigure}{0.32\textwidth}
\centering
\includegraphics[width=\linewidth]{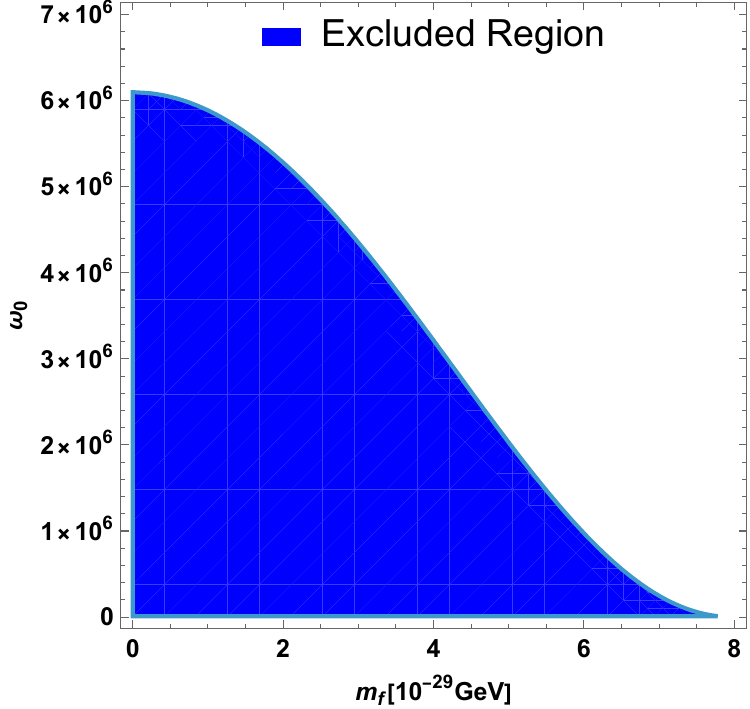}
\caption{}
\label{fig:JVTplot} 
\end{subfigure}
\begin{subfigure}{0.32\linewidth}
 \centering
  \includegraphics[width=\linewidth]{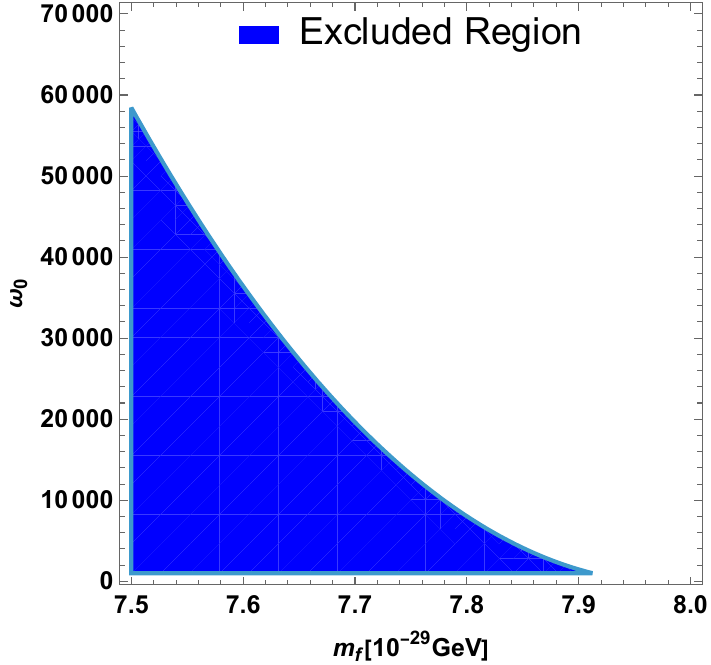}
\caption{}
\label{fig:zoom} 
\end{subfigure}
\caption{(a) Reproduction of Figure 1 in Ref. \cite{Mahmoudi:2024rga}, from where the authors extracted their main result. (b) Corrected bound on $\omega_0$; the new result inverts the excluded region in comparison to (a). (c) Zoom in on figure (b) for the maximum allowed values of $m_f$; it shows the expected recovery of the traditional low mass bound of $\omega_0 \gtrsim 40000$, and how it is weakened for larger mass values.}
\label{fig:1}
\end{figure*}

Our second remark concerns the computation of the aforementioned constraint on $\omega_0$. As stated in equation $(120)$ of Ref. \cite{Mahmoudi:2024rga}, the fractional period decay $\dot{T}$ due to the emission of gravitational wave radiation in the BD$f$ theory given by action \eqref{action} reads
{\small %
\begin{align}
        \dot{T}  = & \, \frac{\dot{T}_{\mathrm{GR}}(1-\xi)}{G^{\,2/3}}\,\tilde{\mathfrak{g}}^{2/3}\,\bigg\{ \tfrac{1}{12}-\tfrac{5}{144}\left[ \tfrac{2\,\pi\, m\,\tilde{\mathfrak{g}}}{T} \right]^{-2/3}\mathcal{S}^{2}\,\times\notag\\[1ex]
        & \,\bigg[ \Big( 1-\tfrac{T^{2}m_{f}^{2}}{4\pi^{2}} \Big)^{^{3/2}}\!\!\!\! \Theta(\omega-m_{f})  -\tfrac{2\omega_{0}+1}{4\omega_{0}^{2}(2\omega_{0}+3)}\,\Big(\tfrac{3(2\omega_{0}+1)}{2} \,+ \notag \\[1ex]
        & \qquad\quad \left(1-\tfrac{T^{2}m_{f}^{2}}{4\pi^{2}} \right)^{^{1/2}} 
        \!\!\!\!\!\cos\left(\tfrac{\mathfrak{R}m_{f}^{2}}{\omega^{2}} \right)\Theta\left( \omega - m_{f} \right)\bigg]\bigg\},\label{TdotGBD}
\end{align}}
where $\dot{T}_{\mathrm{GR}}$ is the corresponding dimensionless GR factor, $\xi=(2\omega_0+4)^{-1}$, $\mathfrak{g}\equiv\mathfrak{g}(s_a,\omega_0,m_f)$ --- being $s_a$ the sensitivities --- and $\Theta$ stands for the Heaviside function.  
%\begin{equation}/
%    \dot{T}_{\mathrm{GR}} = -\frac{192 \pi}{5} \frac{q}{(1 + q)^2} \left( \frac{2 \pi G m}{T} \right)^{5/3}.
%\end{equation}
The BD limit in the above expression is quite clear when taking $m_f \rightarrow \infty$. Only in this limit $\omega_0$ can be interpreted as the original BD parameter. 

The bounds presented in Fig.~1 of Ref \cite{Mahmoudi:2024rga} can not be obtained from \eqref{TdotGBD}. By revisiting the analysis proposed by the authors we have obtained new bounds on $\omega_0$, as a function of the geometric scalar field mass $m_f$, which presents a more consistent scenario. Our result relies on the same data used by the authors, the binary system PSR J1012+5307 (Table 1 in \cite{Mahmoudi:2024rga}). In Fig. \ref{fig:MVplot} we reproduced the authors plot and in Fig. \ref{fig:JVTplot} it is shown our result. 
We have also identified the origin of this disagreement. The result presented in Ref. \cite{Mahmoudi:2024rga} can be reproduced by replacing the exponent factor of $3/2$ in the term $(1-T^2 m^2_f / 4 \pi^2)^{3/2}$ by $-3/2$. The reason behind this is, therefore, a typo in the numerical code that generated Fig. \ref{fig:MVplot}. This also explains the inversion of the graph pattern as shown in Fig. \ref{fig:JVTplot}. %Moreover, we can now consistently discuss the BD limit, when $m_f$ is too large (but still small then $8\times 10^{-29}$ eV, as imposed by the binary system orbital period). We see from Fig. \ref{fig:zoom} that in this range the geometric field is already losing influence

We conclude this note with a corrected bound on the coupling parameter $\omega_0$ as a function of $m_f$ in BD-$f(R)$ theories as shown in panels \ref{fig:JVTplot} and \ref{fig:zoom}. Examples of such bounds are
\begin{align}
        &\omega_0 \gtrsim 6\times 10^{6}\quad {\rm for} \quad m_f \gtrsim 1\times 10^{-29} \,\text{GeV};\label{mainresult}\\[1ex]
        &\omega_0 \gtrsim 40000\quad {\rm for} \quad m_f \gtrsim7.6\times 10^{-29} \,\text{GeV} .\label{trad-result}
\end{align}
The result presented in \eqref{mainresult} refers to the $m_f$ value regime contrary to the one in Ref. \cite{Mahmoudi:2024rga}, while \eqref{trad-result} shows the Cassini-equivalent bound. It is also important to mention that for larger mass values, as the theory approaches the traditional BD limit, the $\omega_0$ bound becomes weaker. However, note that the BD limit cannot be investigated since there is an upper bound limit for $m_f < 8 \times 10^{-29}$ GeV imposed by the PSR J1012+5307 binary system's orbital period.
\vspace*{-.2cm}
\begin{acknowledgements}
We thank S. Mahmoudi for useful correspondence and for sharing the numerical code used in their work. JDT is partially supported by FAPES, grant No. 1020/2022. The authors also thank Brazilian funding agencies CAPES, CNPq and FAPEMIG for support. 
\end{acknowledgements}

\end{document}